\documentclass[epj]{svjour}
\usepackage{graphicx,float}
\usepackage{amssymb,amsmath,amsfonts,latexsym}
\usepackage{bm}
\usepackage{psfrag}
\usepackage{epsfig}
\usepackage{subfigure}
\usepackage{slashed}
\usepackage{cite}

\begin{document}

\title{Fermi surface renormalization and quantum confinement in the two-coupled chains model}
\author{Eberth Correa \inst{1} \and Alvaro Ferraz \inst{2}}
\institute{Faculdade UnB Gama, Universidade de Bras\'ilia - UnB, 72444-240, Gama-DF, Brazil. \and
International Institute of Physics and Dept. of Theoretical and Experimental Physics- UFRN, 1722, 59078-400, Natal-RN, Brazil.} 

\date{Received: \today / Revised version: \today}
%

\abstract{
We address the problem of the Fermi surface renormalization and the quantum confinement regime (QCR) in the two coupled chains model(TCCM) of spinless fermions. We perform a self-consistent calculation of the renormalization group(RG) flows of the renormalized TCCM couplings and quasiparticle weight. On top of that we take explicitly into account the renormalization of the Fermi surface. The flow of the difference of the renormalized Fermi wave vectors associated with the bonding and antibonding bands has a dramatic effect on the single particle spectrum. Although the quasiparticle amplitude is nullified already at intermediate coupling the QCR is only observed at strong coupling. The state associated with this regime has a charge gap and it is not a Luttinger liquid. In contrast, the Fermi liquid regime is stabilized by the umklapp ``$g_2$--like'' interactions at very weak coupling regime.   
}

\titlerunning{Fermi surface renormalization and QCR in the TCCM}
\authorrunning{Eberth Correa \and Alvaro Ferraz}
\maketitle

\section{Introduction}

Low dimensional metals\cite{jerome} are known to be very sensitive to quantum fluctuations. Those fluctuations produce new exotic phases and may lead, in particular, to Fermi surface(FS) reconstruction. Notable examples of that are the cuprates. At optimal doping the FS is large and hole like. In contrast, at low doping the FS is reduced in size and transport experiments detect the presence of electron pockets instead\cite{doiron}. Ever since the discovery of high Tc superconductors there has been great interest in ladder systems as possible prototypes of strongly correlated two--dimensional materials. These models are specially suitable to address the problems of FS reconstruction induced by electron correlation and the onset of non--conventional superconductivity.      

The ladder systems consist of $N$ one dimensional chains coupled together by a transverse interchain hopping $t_{\perp}$. Several systematic investigations have been performed to elucidate the properties of such models. The simplest of these systems is the so--called two coupled chains model (TCCM). This model describes two spinless Tomonaga--Luttinger chains coupled by $t_{\perp}$. The presence of this transverse hopping produces unexpected changes in the Luttinger liquid (LL)\cite{tomonaga,luttinger,haldane,alvaro1,alvaro4} properties of the two isolated chains.  

Contrary to what happens for a single chain the TCCM is not exactly solvable. In the absence of electron--electron interactions it can be diagonalized exactly and mapped into a system of two non--interacting bands. The resulting FS consists of four Fermi points associated with the so--called bonding and antibonding bands. In this way $t_{\perp}$ is directly determined by the difference of the Fermi points(FP), $\Delta k_{F}=k_{F}^b-k_{F}^a$. The presence of interactions may greatly renormalize $\Delta k_{F}$ and this in turn leads to the nullification of $t_{\perp}$ at sufficiently large coupling. This limit is called the quantum confinement regime(QCR). The QCR will be fully discussed in this work.  

The TCCM was extensively studied using either bo\-so\-ni\-za\-tion\cite{giamarchi,parola,kl,kn,sn,shi2} or RG\cite{nickel,tosatti,fabrizio,Dusuel,ledowski,ledowski2,orignac} methods. The first studies considered only some of the scattering processes produced by forward like couplings associated with the interband and the intraband interactions. The results which were brought about by those analysis soon contradicted the expectations that the LL state would be stable for a small non--zero $t_{\perp}$. In fact, the two coupled LL chains are driven to a Fermi liquid(FL) regime at very weak coupling. 

The key for one to arrive at the QCR\cite{anderson} is the strong renormalization of $t_{\perp}$. Early calculations using bosonization(for the referred choice of interactions) predicted the coexistence of a finite $t_{\perp}$ and a strong interchain pair coherence in a sufficiently strong coupling regime with a deconfinement--confinement transition of a Kosterlitz--Thouless nature\cite{kl,kn,nersesyan}. This scenario is severely modified when the interchain backscattering interaction $g_{\mathcal{B}}$ is present in the model\cite{fabrizio}. One finds instead that this kind of interaction can lead to a new strong coupling regime with the suppression of  $t_{\perp}$ been accompanied by the opening of a correlated charge density wave gap. As we will see later on this is in agreement with the findings in this work. Besides the couplings already mentioned we will also take into account the interchain umklapp interaction $g_{\mathcal{U}}$. We assess the main effects produced by these two couplings on the RG flow processes for the main physical quantities of interest. The umklapp interaction also plays an important role to stabilize the LL state at weak--to--intermediate coupling. 

In this work, we report a self-consistent 2-loop RG calculation of all physical quantities associated with the TCCM to test all possible scenarios. We 
determine the RG flows of all the renormalized couplings, of the quasiparticle weight $Z$ and of the renormalized difference of FPs, $\Delta k_{FR}$. In this way we are able to investigate the disappearance of coherent quasiparticle states at intermediate couplings as well as the onset of the QCR in the TCCM at strong coupling. We establish the RG flows for different initial couplings conditions up to 2-loop order and we are able to determine when the renormalized Fermi points indeed merge into each other, leading the TCCM towards a QCR. This QCR only takes place at larger couplings after the LL state suffers a crossover to another NFL state which is no longer metallic and is characterized by a charge gap. As a result in this regime the TCCM is likely to be in the same universality class of the one dimensional Hubbard model near half-filling. The QCR essentially confines the TCCM back to an effective 1D behavior. As it will become clear, the different physical states present in the TCCM model are directly related to two sets of fixed points of the renormalized couplings.   

\begin{figure}[b]
\centering
\includegraphics[scale=0.25]{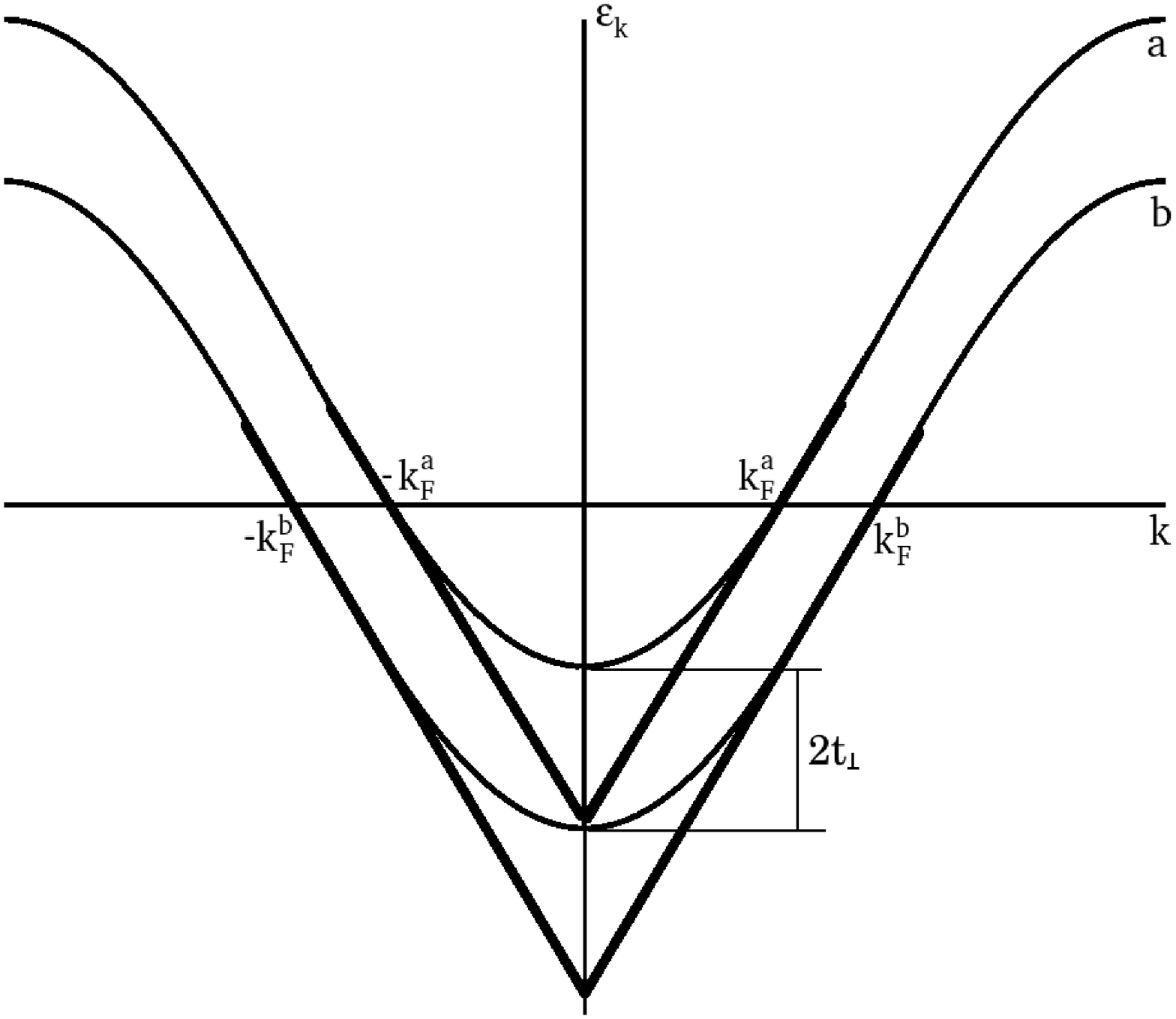}
\caption{Non-interacting single particle dispersions in the TCCM. The antibonding band a, $\varepsilon_{k}^{a}$, and
the bonding band, $\varepsilon_{k}^{b}$.} 
\label{duasbandas}
\end{figure}

\section{The TCCM Model}

The TCCM model basically describes the low--lying excitations of spinless fermions in the presence of two Luttinger liquid chains coupled by a transverse interchain hopping $t_\perp$. The primary effect of this kind of hopping is to split the energy spectrum into two bands as showed in Fig. \ref{duasbandas}. It follows immediatelly from this that the difference of the bonding and antibonding Fermi points (FPs)$(\varepsilon_k^j=0,j=a,b)$ $k_{F}^{b}-k_{F}^{a}=\Delta k_{F}$ is proportional to the transverse hopping $t_{\perp}$\cite{ledowski}. Therefore, it is appropriate to consider scattering processes which become relevant when $\Delta k_{F}$ is sufficiently small. 

Earlier works, using bosonization methods, estimated the effect produced by $t_{\perp}$ in a different way\cite{giamarchi,kl,kn}. Their approach considered from the beginning the limit in which $t_{\perp}$ is much smaller than either the intrachain hopping or intrachain energy. In contrast with that, in our approach, $t_{\perp}$ is implicitly included in the bonding and antibonding Fermi points since $k_{F}^a=k_F-t_{\perp}/v_{F}$ and $k_{F}^b=k_F+t_{\perp}/v_{F}$.

The TCCM with linearized single--particle dispersions can be written in terms of the fermionic fields $\psi_{\alpha}^{j}({\bf{k}})$, where ${\bf{k}}=(k_{0},k)$, $j=a,b$ is the band index (flavor index), with the indices $a$ and $b$ refering to the antibonding and bonding bands, respectively. Here $\alpha$ is the chirality index, representing the right and left-hand sides of the FPs. In this way, the TCCM is fully characterized by two ``flavors" ($a$ and $b$) and two chiralities representing the left ($-$) and right ($+$) handed spinless fermions. Following all these considerations, the TCCM non-interacting Lagrangian density is given by

\begin{eqnarray}
\mathcal{L}_{0}({\bf{k}}) = \sum_{\alpha,j}\psi^{j\dagger}_{\alpha}({\bf{k}})\left(k_{0} 
-\varepsilon_{\alpha k}^{j}\right)\psi_{\alpha}^{j}({\bf{k}}), 
\label{lzero}
\end{eqnarray}

\noindent where $k_{0}$ is the single particle frequency and $\varepsilon_{\alpha k}^{j}=v_{F}\big(\alpha k -k_{F}^{j}\big)$ is the linearized single-particle 
energy dispersion around the $\pm k_{F}^{j}$ FPs.

\begin{figure}[t]
\centering
\includegraphics[scale=0.50]{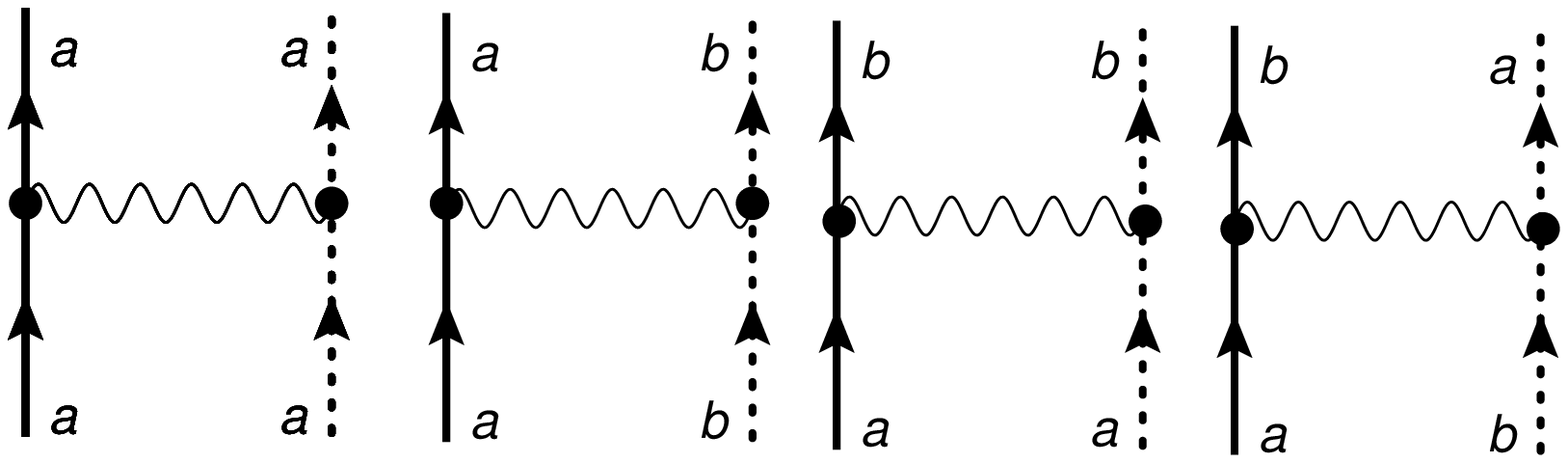}
\caption{Band interactions in the TCCM: intra--forward $-ig_{0,R}$, inter--forward $-ig_{\mathcal{F},R}$, 
umklapp $-ig_{\mathcal{U},R}$ and backscattering $-ig_{\mathcal{B},R}$. The solid(dashed) line refers to the right(left) electrons with respect to the Fermi point $k_{F}^{j}(-k_{F}^{j})$.}
\label{interacoesTCM}
\end{figure}

\begin{figure}[b]
\centering
\includegraphics[scale=0.5]{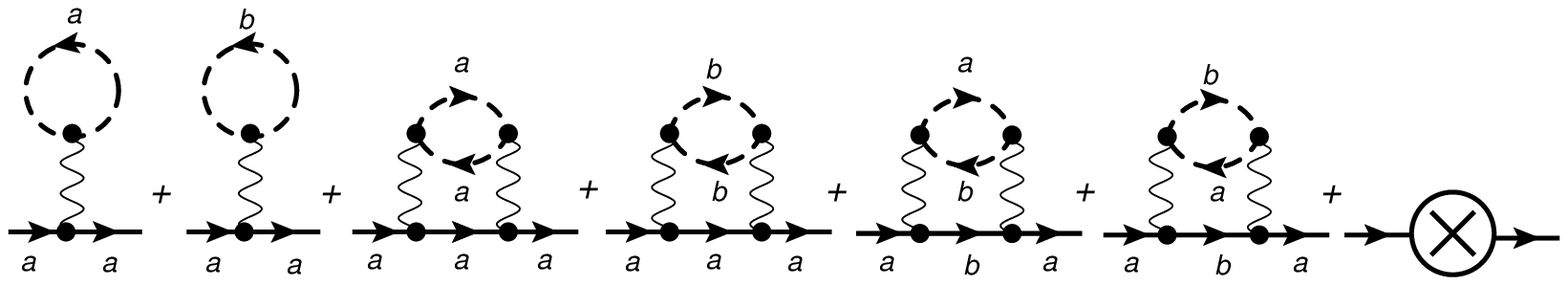}
\caption{$G_{+R}^{a(2)}$ diagrams up to 2-loops.}
\label{selfenergyb}
\end{figure}

Here we consider the interactions\cite{fabrizio} which represent all possible intra and inter chain scatterings taking explicitly into account only the forward like interaction ex\-chan\-ges between particles in the bonding and antibonding bands as shown in Fig. \ref{interacoesTCM}. Using the convention that the label a(b) ``flavor'' refers to the antibonding (bonding) particles we display all these interactions explicitly in Fig. \ref{interacoesTCM}. These interactions preserve chiral symmetry although flavor symmetry is explicitly violated by the umklapp coupling.

The interacting TCCM action is therefore given by
\begin{eqnarray}
\mathcal{S}_{\texttt{int}}&=&-\sum_{\lbrace {\bf{k}}_{i}\rbrace,\alpha,j}\lbrace 
g_{0}\hat{\psi}^{j\dagger}_{\alpha}({\bf{k}}_{3})
\hat{\psi}^{j\dagger}_{-\alpha}({\bf{k}}_{4})
\hat{\psi}_{\alpha}^{j}({\bf{k}}_{2})\hat{\psi}_{-\alpha}^{j}({\bf{k}}_{1}) 
\nonumber \\
&+&g_{\mathcal{F}}\hat{\psi}^{j\dagger}_{\alpha}({\bf{k}}_{3})\hat{\psi}^{-j\dagger}_{-\alpha}({\bf{k}}_{4})
\hat{\psi}_{\alpha}^{j}({\bf{k}}_{2})\hat{\psi}_{-\alpha}^{-j}({\bf{k}}_{1}) \nonumber \\
&+&g_{\mathcal{B}}\hat{\psi}^{j\dagger}_{\alpha}({\bf{k}}_{3})\hat{\psi}^{-j\dagger}_{-\alpha}({\bf{k}}_{4})
\hat{\psi}_{\alpha}^{-j}({\bf{k}}_{2})\hat{\psi}_{-\alpha}^{j}({\bf{k}}_{1}) \nonumber \\
&+&g_{\mathcal{U}}\hat{\psi}^{j\dagger}_{\alpha}({\bf{k}}_{3})\hat{\psi}^{j\dagger}_{-\alpha}({\bf{k}}_{4})
\hat{\psi}_{\alpha}^{-j}({\bf{k}}_{2})\hat{\psi}_{-\alpha}^{-j}({\bf{k}}_{1}) \rbrace, 
\nonumber \\ 
\end{eqnarray}

\begin{figure}[t]
\centering
\subfigure[Interband forward channel.]{\includegraphics[scale=0.37]{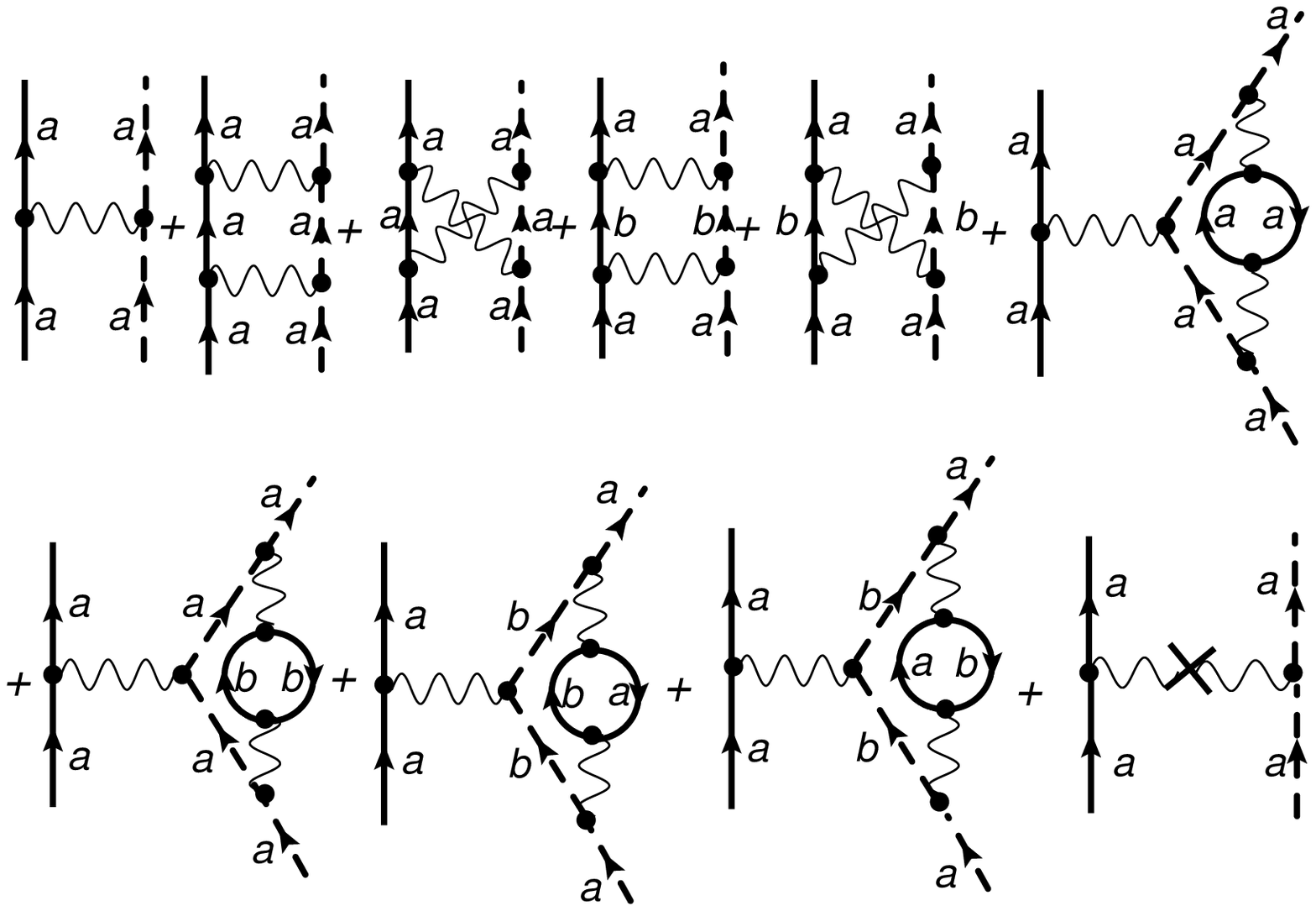}
}
\subfigure[Forward channel.]{\includegraphics[scale=0.37]{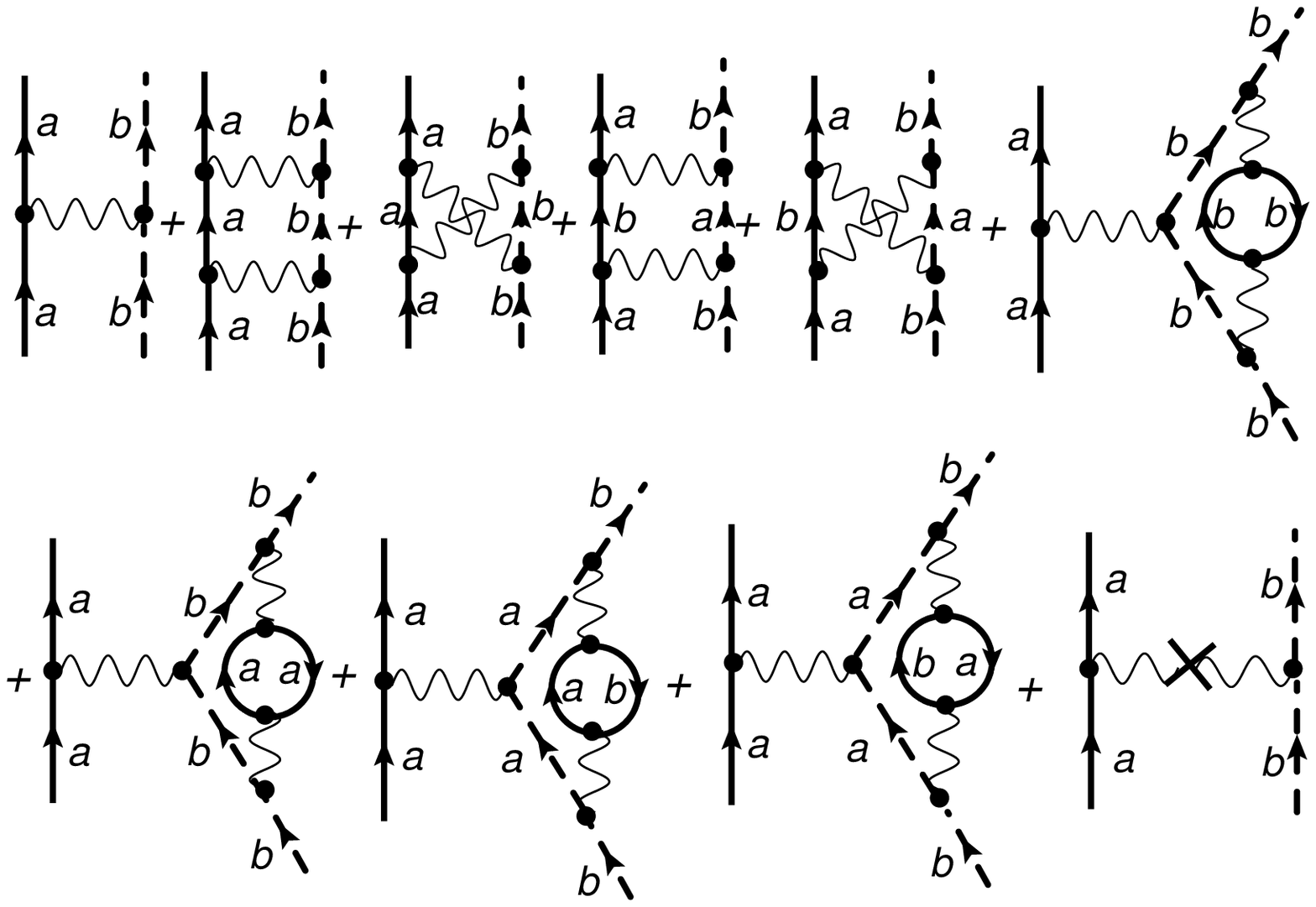}
}
\caption{$\Gamma^{(4)}_{iR}(\lbrace{\bf{k}}_{m}\rbrace)$ diagrams up to 2-loops, for $i=0,\,\mathcal{F}$.}
\label{gamma4}
\end{figure}

\noindent where ${\bf{k}}_{4}={\bf{k}}_{1}+{\bf{k}}_{2}-{\bf{k}}_{3}$ by energy--momentum conservation, $\{\mathbf{k}_i\}=\mathbf{k}_1,\,\mathbf{k}_2,\,\mathbf{k}_3$ and we use the convention that $-j$  refers to the other $j$ ``flavor'', \emph{i.e.}, $-b=a$ and $-a=b$. 

Notice that as emphasized before all couplings in the interacting TCCM action above are forward like with respect to their chirality and they only differ from each other by their flavor characteristics. Thus, one must avoid mistaking the interband backscattering and umklapp interactions considered here with the corresponding backscattering and umklapp couplings from the single chain model. In other words, basically, all TCCM interactions are ``$g_2$--like'' and they are associated with small energy--momentum transfers only. The corresponding $g_4$ interactions are not considered here, for convenience, in view of the fact that they only contribute one order in perturbation theory late (in the infrared limit) in comparison with the other forward like couplings.  

\begin{figure}[t]
\centering
\subfigure[Backscattering channel.]{\includegraphics[scale=0.37]{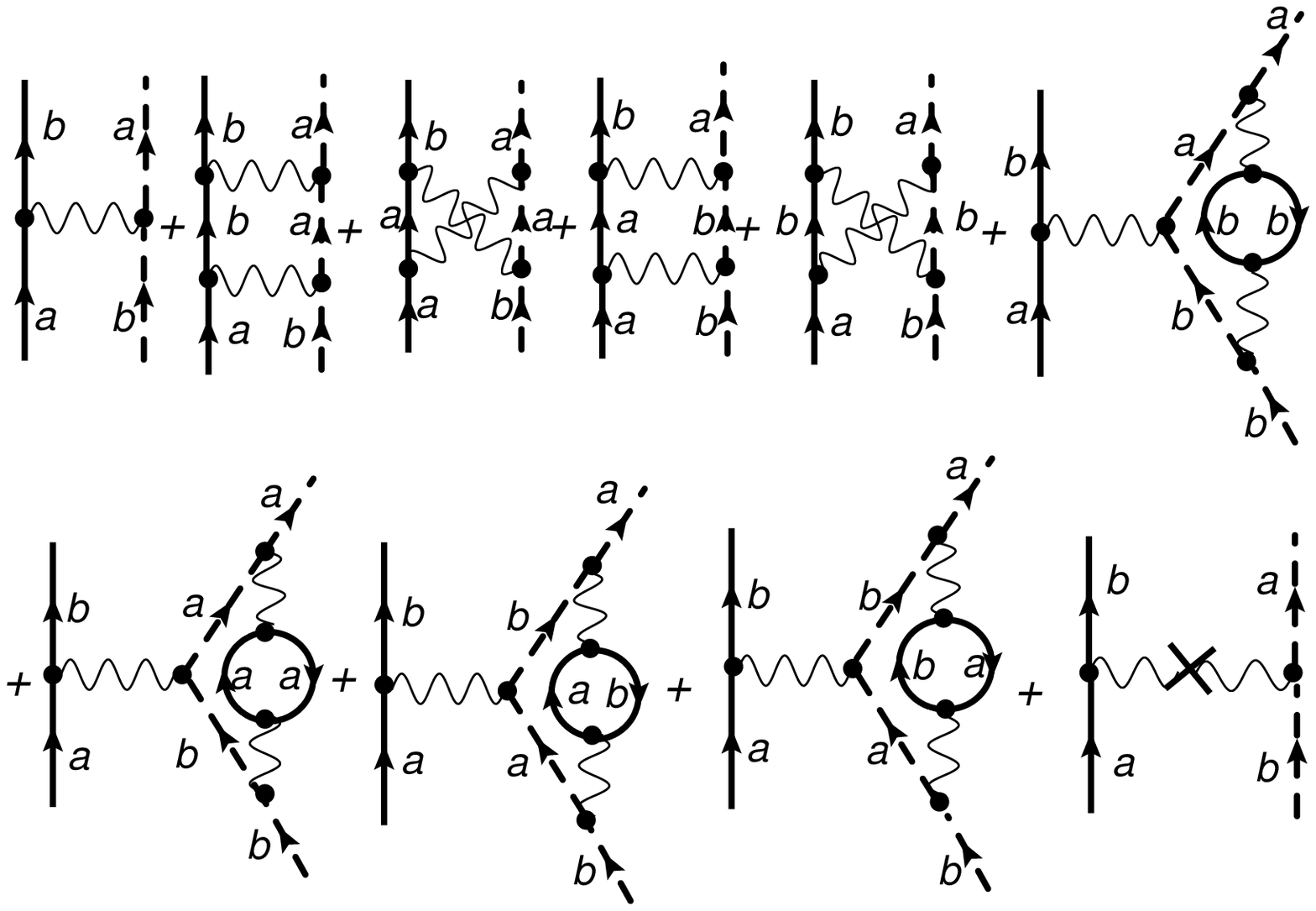}
}
\subfigure[Umklapp channel.]{\includegraphics[scale=0.37]{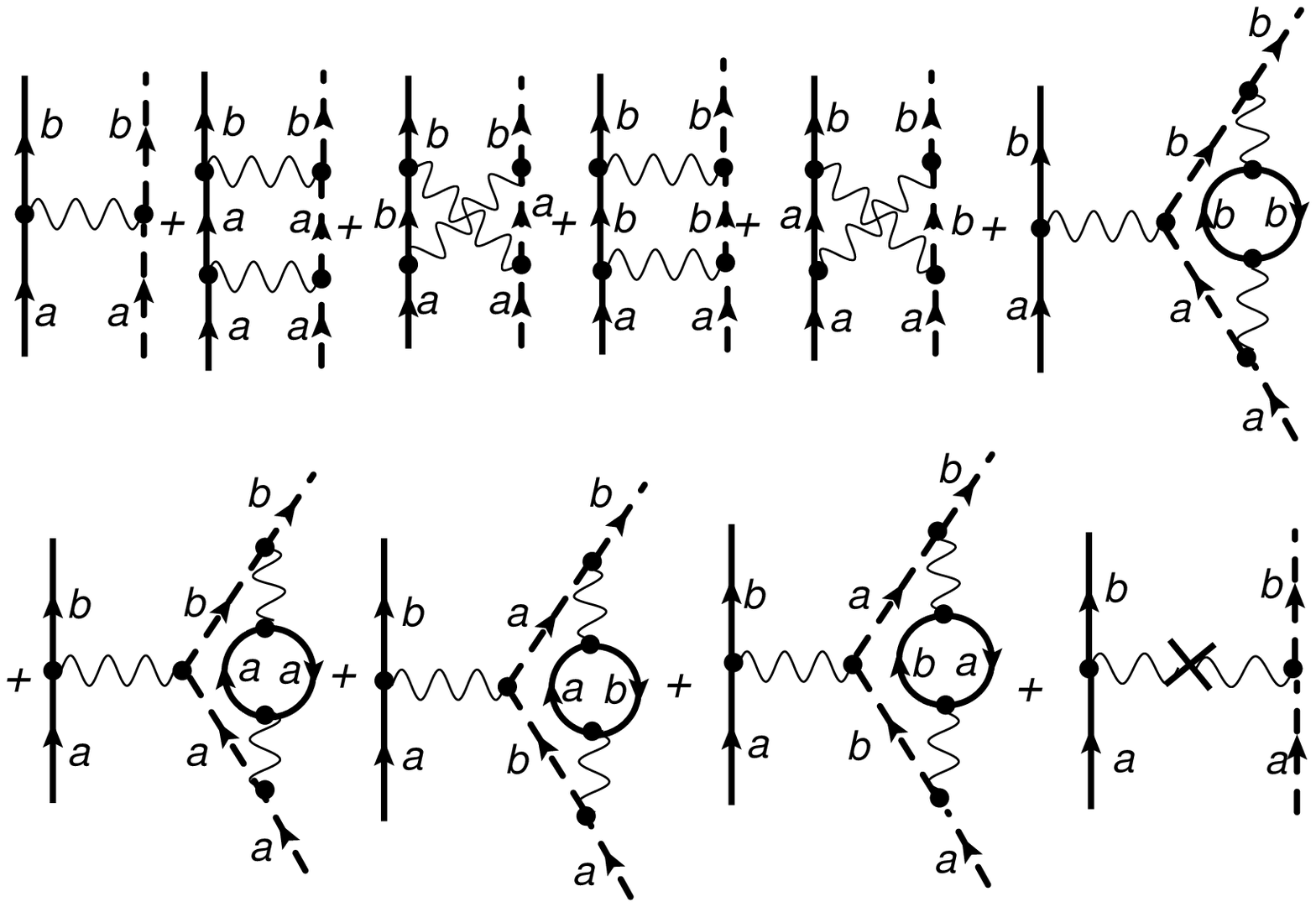}
}
\caption{$\Gamma^{(4)}_{iR}(\lbrace{\bf{k}}_{m}\rbrace)$ diagrams up to 2-loops, for $i=\mathcal{B},\,\mathcal{U}$.}
\label{gamma4b}
\end{figure}

\section{The Field Theoretical Renormalization Group Treatment}

In view of the linear single--particle dispersions and in virtue of the Dirac seas for each of the ``$a$'' and ``$b$'' bands the resulting TCCM action produces both ultraviolet (UV) and infrared (IR) divergences. The UV divergencies manifest themselves already in one-loop order and are directly associated with the chiral anomalies which enforce the definition of a generalized Ward-Takahashi identity (WTI) which is discussed elsewhere \cite{alvaro2}. For the calculation performed here we follow closely the field theoretical renormalization group treatment\cite{peskin}. As a result we make use of proper renormalization prescriptions to determine explicitly the counterterms which are needed to render the theory free from divergences order by order in perturbation theory\cite{freire1}. In this way, the renormalized parameters are defined in accordance with the prescribed values of the related renormalized one-particle irreducible functions at the Fermi points(FPs). Here, we choose conventional renormalization prescriptions taking into consideration that the ``bare'' quantities do not know anything about the RG energy scale $\omega$. Consequently, in our RG scheme the re\-nor\-ma\-lized one--particle irreducible functions $\Gamma^{j (2)}_{\alpha R}$ and $\Gamma_{iR}^{(4)}$ are determined by the conditions

\begin{eqnarray}
\Gamma^{j (2)}_{\alpha R}(\omega,p)\vert_{FP} &=& \left(G^{j(2)}_{\alpha R}({\bf{k}})\right)^{-1}_{FP}=\omega \label{prescriptionsa} \\
\frac{\partial\Gamma^{j (2)}_{\alpha R}(\omega,p)}{\partial p}\vert_{FP} &=& \pm v_{FR} \label{prescriptionsb}  \\
\Gamma_{iR}^{(4)}(\lbrace {\bf{k}}_{m} \rbrace)\vert_{FP} &=& -ig_{iR},  \label{prescriptionsc}
\end{eqnarray}

\noindent where $FP$ stands for the related Fermi point. In this way all renormalized physical quantities, which are functions of energy and momentum, are calculated at the Fermi points. 

Following the field theoretical RG procedure\cite{peskin,freire1,freire4} to construct the renormalized action we simply add an appropriate counterterm for each corresponding term of the ``bare'' action. It then follows that the renormalized TCCM noninteracting Lagrangian density can be written as\cite{peskin}
\begin{eqnarray}
\mathcal{L}_{0R}({\bf{k}})&=&\sum_{\alpha,j}\lbrace \psi^{j\dagger}_{\alpha R}({\bf{k}})\left[k_{0} 
- v_{FR}\left(\alpha k -k_{FR}^{j}\right)\right]\psi^{j}_{\alpha R}({\bf{k}}) \nonumber \\
&+& \psi^{j\dagger}_{\alpha R}({\bf{k}})\left[ k_{0}\delta Z 
-\alpha k\delta v_{FR} + v_{FR}\delta k_{F}^{j}\right]\psi^{j}_{\alpha R}({\bf{k}})\rbrace \nonumber \\
\end{eqnarray}
where $\delta Z=Z-1$, with $Z$ being the quasiparticle weight, $\delta v_{FR}^{j}=(ZZ_{v_{FR}^{j}}-1)v_{FR}^{j}$ 
and $\delta k_{FR}^{j}=(ZZ_{v_{FR}^{j}}Z_{k_{FR}^{j}}-1)k_{FR}^{j}$, with $Z_{v_{FR}^{j}}$ 
and $Z_{k_{FR}^{j}}$ being the corresponding form factors for the renormalized Fermi velocity and the renormalized Fermi 
wavevector. 

We proceed in the same way with the renormalized interacting action and as a result we find that 

{\small
\begin{eqnarray}
\begin{split}
\mathcal{S}_{\texttt{int}}^R=-\sum_{\lbrace {\bf{k}}_{i}\rbrace,\alpha,j}&\lbrace 
\left(g_{0R} + \delta g_{0R}\right)\times \nonumber \\ 
&\hat{\psi}^{j\dagger}_{\alpha R}({\bf{k}}_{3})
\hat{\psi}^{j\dagger}_{-\alpha R}({\bf{k}}_{4})
\hat{\psi}_{\alpha R}^{j}({\bf{k}}_{2})\hat{\psi}_{-\alpha R}^{j}({\bf{k}}_{1}) 
\end{split}\nonumber \\
\begin{split}
&+\left(g_{\mathcal{F}R} + \delta g_{\mathcal{F}R}\right)
\hat{\psi}^{j\dagger}_{\alpha R}({\bf{k}}_{3})\hat{\psi}^{-j\dagger}_{-\alpha R}({\bf{k}}_{4})
\hat{\psi}_{\alpha R}^{j}({\bf{k}}_{2})\hat{\psi}_{-\alpha R}^{-j}({\bf{k}}_{1}) \nonumber \\
&+\left(g_{\mathcal{B}R} + \delta g_{\mathcal{B}R}\right)  
\hat{\psi}^{j\dagger}_{\alpha R}({\bf{k}}_{3})\hat{\psi}^{-j\dagger}_{-\alpha R}({\bf{k}}_{4})
\hat{\psi}_{\alpha R}^{-j}({\bf{k}}_{2})\hat{\psi}_{-\alpha R}^{j}({\bf{k}}_{1}) \nonumber \\
&+\left(g_{\mathcal{U}R} + \delta g_{\mathcal{U}R}\right)  
\hat{\psi}^{j\dagger}_{\alpha R}({\bf{k}}_{3})\hat{\psi}^{j\dagger}_{-\alpha R}({\bf{k}}_{4})
\hat{\psi}_{\alpha R}^{-j}({\bf{k}}_{2})\hat{\psi}_{-\alpha R}^{-j}({\bf{k}}_{1}) \rbrace, 
\end{split}\nonumber \\
\end{eqnarray} 
}

\noindent where ${\bf{k}}_{4}={\bf{k}}_{1}+{\bf{k}}_{2}-{\bf{k}}_{3}$ by energy--momentum conservation and $\delta g_{0R}$, $\delta g_{\mathcal{F}R}$, $\delta g_{\mathcal{B}R}$ and $\delta g_{\mathcal{U}R}$ are the accompanying counterterms 
for the renormalized couplings. In Fig. \ref{selfenergyb} we display the diagrams for the renormalized fermionic single-particle propagator $G^{j(2)}_{\alpha R}({\bf{k}})$. In Figs. \ref{gamma4} and \ref{gamma4b} we display the diagrams for the renormalized one particle irreducible four-point functions, in the ``$i$-th'' scattering channel 
( $i=0,\mathcal{F},\mathcal{B},\mathcal{U}$) $\Gamma^{\alpha(4)}_{i,R}(\lbrace{\bf{k}}_{i}\rbrace)$ up to 2-loop order.

Using the Feynman rules associated with the renormalized TCCM Lagrangian density $\mathcal{L}_{R}$ we calculate all $\Gamma^{(2)}_{R}$'s as well 
as the four $\Gamma^{(4)}_{iR}$'s of interest up to 2-loop order as displayed in Figs. \ref{selfenergyb}, \ref{gamma4} and  \ref{gamma4b}. For convenience, we introduce the ultraviolet cutoff $\Lambda_0$ in the momentum space such that $\Omega=2v_{FR}\Lambda_0>\omega$ and we find that

\begin{align}
&\delta Z=-\frac{1}{2}[\bar{g}_{\mathcal{F}R}^2+\bar{g}_{0R}^2+\bar{g}_{\mathcal{U}R}^2]\ln
\bigg(\frac{\Omega}{\omega}\bigg)-\frac{\bar{g}_{\mathcal{B}R}^2}{4}\times \nonumber \\ &\bigg[\ln\bigg(\frac{\Omega}{2v_{FR}\Delta k_{FR}+\omega}\bigg)+\ln\bigg(\frac{\Omega}{|2v_{FR}\Delta k_{FR}-\omega|}\bigg)\bigg]\label{deltaz}  
\end{align}

\noindent and

\begin{eqnarray}
\delta v_{FR}&=&v_{FR}\delta Z \label{deltavf} \\
\delta k_{FR}^{a}&=&k_{FR}^{a}\delta Z - \frac{1}{2}\Delta k_{FR}\bar{g}_{\mathcal{B}R}^{2}\bigg[\ln\bigg(\frac{\Omega}{2v_{FR}\Delta k_{FR}+\omega}\bigg)\nonumber \\
&+&\ln\bigg(\frac{\Omega}{|2v_{FR}\Delta k_{FR}-\omega|}\bigg)\bigg] \label{deltakfa} \\
\delta k_{FR}^{b}&=&k_{FR}^{b}\delta Z
+ \frac{1}{2}\Delta k_{FR}\bar{g}_{\mathcal{B}R}^{2}\bigg[\ln\bigg(\frac{\Omega}{2v_{FR}\Delta k_{FR}+\omega}\bigg)\nonumber \\
&+&\ln\bigg(\frac{\Omega}{|2v_{FR}\Delta k_{FR}-\omega|}\bigg)\bigg] \label{deltakfb}
\end{eqnarray}

\noindent where $\bar{g}_{i}=g_{i}/2\pi v_{FR}$ is now a dimensionless quantity. The IR parameter $\omega$ is our RG energy scale. However, from the numerical point of view, it is extremely convenient to rewrite our RG scale parameter in terms of the variable $l$ which grows indefinitely as $\omega$ flows to zero. We can identify $\Omega$ with a fixed upper cutoff and both variables are related to each other if we take $\omega=\Omega e^{-l}$ and, in this way, $\omega\rightarrow0$ when $l\rightarrow\infty$\cite{freire1}.  

By considering the prescriptions (\ref{prescriptionsa}), (\ref{prescriptionsb}) and (\ref{prescriptionsc}) it follows immediately  that $Z_{v_{FR}}=1$. As a result $v_{FR}$ suffers no independent renormalization. On the other hand, the same is not true for the $k_{FR}^{j}$'s.  

If we take explicitly into account the no flow condition of the bare parameters with respect to $\omega$(or $l$) we readily arrive at the 
2-loop RG flow equations 

{\small
\begin{eqnarray}
\frac{d\bar{g}_{\mathcal{B}R}(l)}{dl}=&-&2\gamma(l)\bar{g}_{\mathcal{B}R}-\bar{g}_{\mathcal{B}R}\left[
\bar{g}_{\mathcal{F}R} -\bar{g}_{0R}-\bar{g}_{\mathcal{U}R}^2\right]\nonumber \\
&-&\frac{\bar{g}_{\mathcal{B}R}}{2}\left(\bar{g}_{\mathcal{F}R}-\bar{g}_{0R}+2\bar{g}_{0R}
\bar{g}_{\mathcal{F}R}-\bar{g}_{\mathcal{U}R}^2\right)\nonumber \\
&\times &\left[\frac{1}{1+\frac{\Delta k_{FR}}{\Lambda_0}e^l}+\frac{sign(1-\frac{\Delta k_{FR}}{\Lambda_0}e^l)}{\left|1-\frac{\Delta k_{FR}}{\Lambda_0}e^l\right|}\right] \label{rgeq1}\\
\frac{d\bar{g}_{0R}(l)}{dl}=&-&2\gamma(l)\bar{g}_{0R}-\bar{g}_{\mathcal{U}R}^{2}+\bar{g}_{0R}
\left(\bar{g}_{0R}^2+\bar{g}_{\mathcal{F}R}^{2}\right)\nonumber \\
&+&\bar{g}_{\mathcal{F}R}\bar{g}_{\mathcal{U}R}^{2} +
\frac{\bar{g}_{\mathcal{B}R}}{2}\left(\bar{g}_{\mathcal{B}R}+\bar{g}_{\mathcal{F}R}
\bar{g}_{\mathcal{B}R}\right)\nonumber \\
&\times &\left[\frac{1}{1+\frac{\Delta k_{FR}}{\Lambda_0}e^l}+\frac{sign(1-\frac{\Delta k_{FR}}{\Lambda_0}e^l)}{\left|1-\frac{\Delta k_{FR}}{\Lambda_0}e^l\right|}\right]\label{rgeq2}
\end{eqnarray}
}
{\small
\begin{eqnarray}
\frac{d\bar{g}_{\mathcal{F}R}(l)}{dl}=&-&2\gamma(l)\bar{g}_{\mathcal{F}R}+\bar{g}_{\mathcal{U}R}^2 +\bar{g}_{\mathcal{F}R}\left(
\bar{g}_{0R}^2+\bar{g}_{\mathcal{F}R}^2\right)\nonumber \\
&+&\bar{g}_{0R}\bar{g}_{\mathcal{U}R}^2
-\frac{1}{2}\left(\bar{g}_{\mathcal{B}R}^2-\bar{g}_{0R}\bar{g}_{\mathcal{B}R}^2\right)\nonumber \\&\times &\left[\frac{1}{1+\frac{\Delta k_{FR}}{\Lambda_0}e^l}+\frac{sign(1-\frac{\Delta k_{FR}}{\Lambda_0}e^l)}{\left|1-\frac{\Delta k_{FR}}{\Lambda_0}e^l\right|}\right] \label{rgeq3}\\
\frac{d\bar{g}_{\mathcal{U}R}(l)}{dl}=&-&2\gamma(l)\bar{g}_{\mathcal{U}R}-2\bar{g}_{\mathcal{U}R}\bar{g}_{0R}+
2\bar{g}_{\mathcal{U}R}\bar{g}_{\mathcal{F}R}\nonumber \\&+&2\bar{g}_{\mathcal{U}R}\bar{g}_{0R}\bar{g}_{\mathcal{F}R}+
\bar{g}_{\mathcal{U}R}\bar{g}_{\mathcal{B}R}^2\nonumber \\
&\times &\left[\frac{1}{1+\frac{\Delta k_{FR}}{\Lambda_0}e^l}+\frac{sign(1-\frac{\Delta k_{FR}}{\Lambda_0}e^l)}{\left|1-\frac{\Delta k_{FR}}{\Lambda_0}e^l\right|}\right] \label{rgeq4}\\
\frac{d\Delta k_{FR}(l)}{dl}=&-&\bar{g}_{\mathcal{B}R}^2\Delta k_{FR}\nonumber \\
&\times &\left[\frac{1}{1+\frac{\Delta k_{FR}}{\Lambda_0}e^l}+\frac{sign(1-\frac{\Delta k_{FR}}{\Lambda_0}e^l)}{\left|1-\frac{\Delta k_{FR}}{\Lambda_0}e^l\right|}\right] \label{rgeq5}\\
\frac{dZ(l)}{dl}=&-&\gamma(l) Z(l),\label{rgeq6}
\end{eqnarray} 
}
with
{\small
\begin{eqnarray}
\gamma(l) &=& \frac{1}{2}\left(\bar{g}_{0R}^2+\bar{g}_{\mathcal{F}R}^2 +
\bar{g}_{\mathcal{U}R}^2\right)\nonumber \\
&+&\frac{\bar{g}_{\mathcal{B}R}^2}{4}
\left[\frac{1}{1+\frac{\Delta k_{FR}}{\Lambda_0}e^l}+\frac{sign(1-\frac{\Delta k_{FR}}{\Lambda_0}e^l)}{\left|1-\frac{\Delta k_{FR}}{\Lambda_0}e^l\right|}\right]\label{rggama}
\end{eqnarray}
}

Notice that if the $\bar{g}_{iR}$'s approach fixed point values, $\gamma(l)\rightarrow \gamma^{*}$, the anomalous dimension, and in this case we necessarily have $Z(l)\rightarrow 0$, when we take $l \rightarrow \infty$.

In the numerical evolution of these RG flow equations we have to give special attention to the terms in square brackets in Eqs. (\ref{rgeq1})--(\ref{rggama}), since the modulus contribution appearing in the denominator can generate unphysical singularities throughout the RG process. In as much as our RG flows are regulated by the presence of fixed points, we can avoid such a problem making use of a suitable logarithmic approximation\cite{Dusuel} in equations (\ref{deltaz})--(\ref{deltakfb}). That is, we simply rewrite the logarithms inside the square brackets of these equations in terms of the series expansion $\ln{(1+x)}+\ln{(1-x)}= -x^2-\frac{x^4}{2}\,...$ and we arrive at the result

\begin{eqnarray}
\left\{\begin{array}{lr}
2\ln{\left(\frac{\Omega}{2v_{FR}\Delta k_{FR}}\right)}+\left(\frac{\omega}{2v_{FR}\Delta k_{FR}}\right)^2 & \mbox{if $\omega<2v_{FR}\Delta k_{FR}$;} \vspace{0.4cm} \\
2\ln{\left(\frac{\Omega}{\omega}\right)}+\left(\frac{2v_{FR}\Delta k_{FR}}{\omega}\right)^2 & \mbox{if $\omega>2v_{FR}\Delta k_{FR}$.} 
\end{array}\right.
\end{eqnarray}

Now if we derive the equations above bearing in mind the 2--loop order nature of our perturbative theory, as well as the fact that $\omega=\Omega e^{-l}$,  we can simply replace the square brackets in Eqs. (\ref{rgeq1})--(\ref{rggama}) by

\begin{eqnarray}
\left\{\begin{array}{lr}
-2\left(\frac{\Lambda_0}{\Delta k_{FR}}e^{-l}\right)^2 & \mbox{if $\frac{\Delta k_{FR}}{\Lambda_0}e^l>1$;}\vspace{0.4cm} \\
2\left[1+\left(\frac{\Delta k_{FR}}{\Lambda_0}e^l\right)^2\right] & \mbox{if $\frac{\Delta k_{FR}}{\Lambda_0}e^l<1$.} 
\end{array}\right. 
\end{eqnarray}

\begin{figure}[t]
\centering
\subfigure[]{\includegraphics[scale=0.32,angle=270]{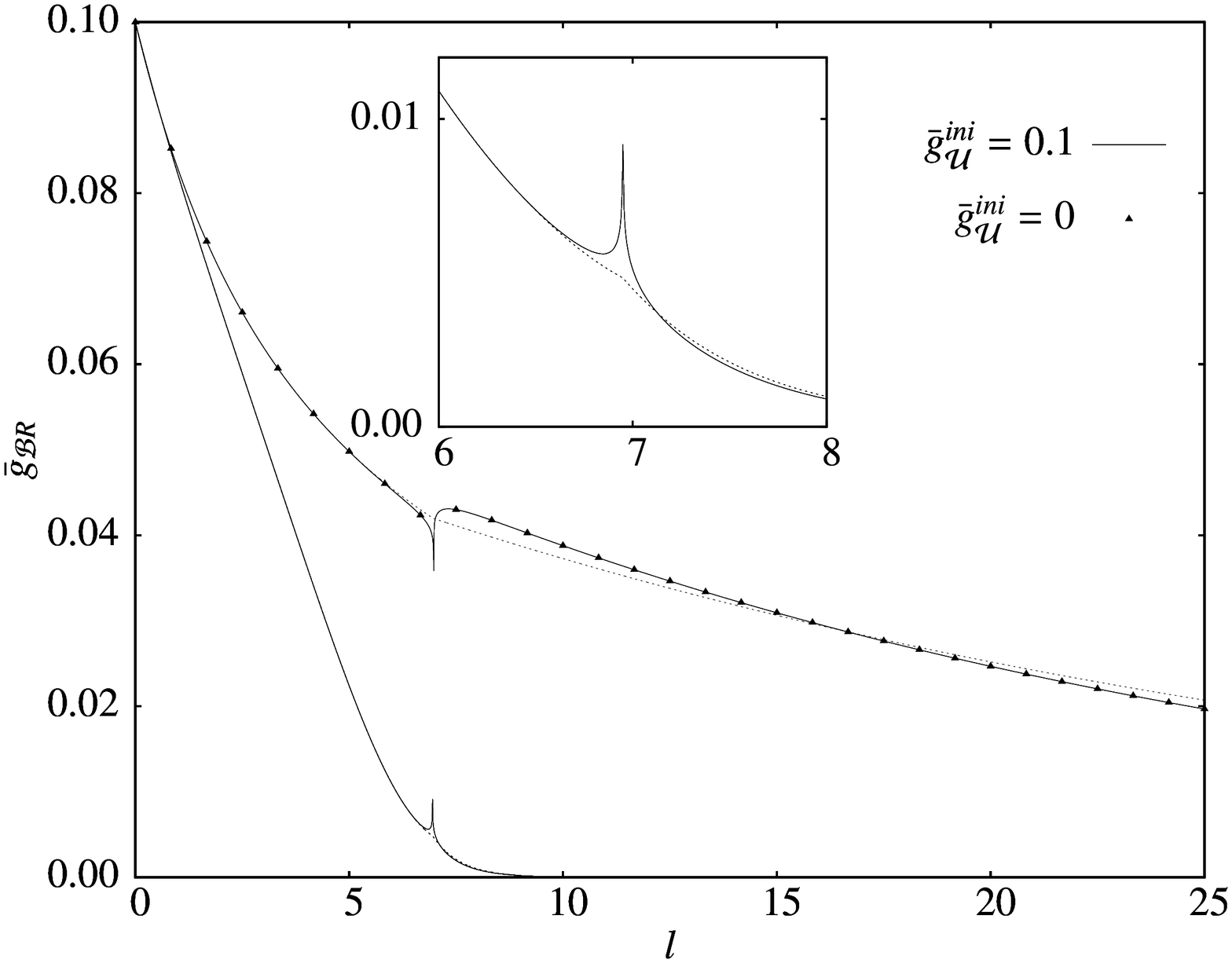}}
\subfigure[]{\includegraphics[scale=0.32,angle=270]{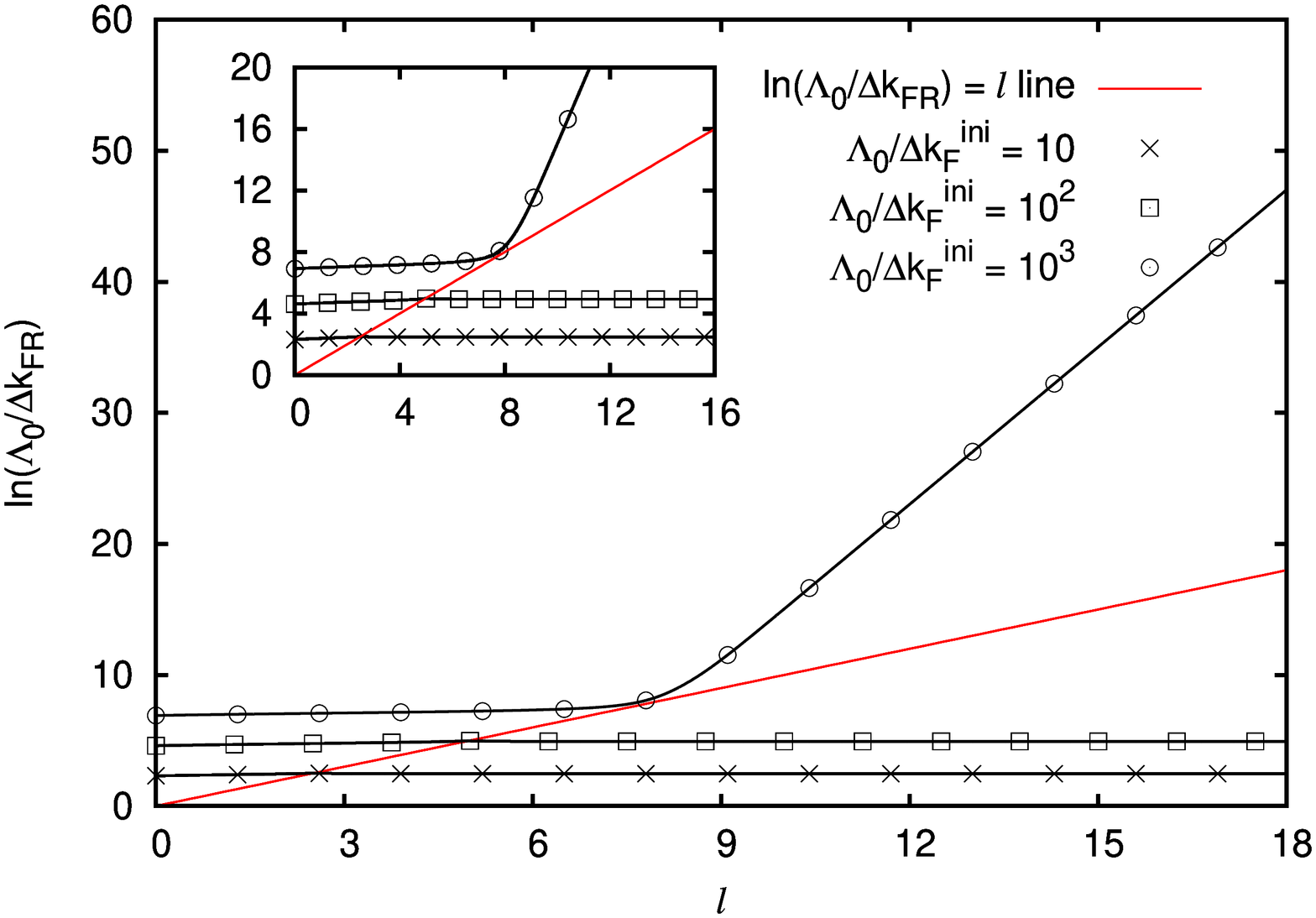}}
\caption{(Color online)(a)The $\bar{g}_{\mathcal{B}R}$ flow with the logarithmic approximation(dashed line) and its full flow without such an approximation(the solid line) for $\bar{g}_{\mathcal{B}}^{ini}=0.1$, $\bar{g}_{0}^{ini}-\bar{g}_{\mathcal{F}}^{ini}=-0.1$, $\bar{g}_{\mathcal{U}}^{ini}=0,\,0.1$ and $\Lambda_0/\Delta k_{F}^ {ini}=10^3$ and (b) $\ln\left(\Lambda_0/\Delta k_{FR}\right)$ versus the RG step $l$ for $\bar{g}_{\mathcal{B}}^{ini}\approx0.212$, $\bar{g}_{0}^{ini}-\bar{g}_{\mathcal{F}}^{ini}=-0.1$ and $\bar{g}_{\mathcal{U}}^{ini}=0.1$ as initial conditions.}
\label{lamb0}
\end{figure}

We display the RG flows for $\bar{g}_{\mathcal{B}R}$ for a specific choice of the initial conditions in Fig. \ref{lamb0}(a). The dashed line represents the flow for $\bar{g}_{\mathcal{B}R}$ with our logarithmic approximation while the solid line displays the related full flow without any approximation. Notice the trend of $\bar{g}_{\mathcal{B}R}$ for different choices of $\bar{g}_{\mathcal{U}}^{ini}$. For $\bar{g}_{\mathcal{U}}^{ini}=0$ the $\bar{g}_{\mathcal{B}R}$ flow never seems to stabilize for our choice of $l_{max}=25$ in the weak coupling regime. Moreover, for a given RG step $l$ value, $\bar{g}_{\mathcal{B}R}$ plays no effect on the Fermi points and $\Delta k_{FR}$ simply stops to be renormalized further in spite of the $\bar{g}_{\mathcal{B}R}$ slow variations. The same happens with the other couplings. This behavior is drastically modified as soon as $\bar{g}_{\mathcal{U}}^{ini}$ takes non--zero values. As we will see, this is directly associated with the onset of the LL state in the weak--to--intermediate coupling regime.   

In Fig. \ref{lamb0}(b) we exhibit $\ln\left(\Lambda_0/\Delta k_{FR}\right)$ as a function of the RG step $l$, for a particular set of initial conditions. Essentially, these initial conditions are determined by our choices of couplings and $\Lambda_0/\Delta k_{F}^{ini}$. It is important to emphasize that this choice is made in such a way that we can keep track of the strong renormalization of $\Delta k_{FR}$ whenever this is the case. In this work the $\Delta k_{FR}$ flow is such that $\ln\left(\Lambda_0/\Delta k_{FR}\right)$ remains only at one side of the line $\ln\left(\Lambda_0/\Delta k_{FR}\right)=l$ as shown in Fig. \ref{lamb0}(b). If the renormalization of $\Delta k_{FR}$ is not strong enough to interfere with all other RG equations, it follows that $\ln\left(\Lambda_0/\Delta k_{FR}\right)$ is not changed significantly and the logarithmic approximation ensures the continuity of the flow throughout the whole RG process, as can be observed in the other two curves in Fig. \ref{lamb0}(b).

As it will be clear next there are two sets of fixed points for the TCCM renormalized couplings. The quantum confinement regime is connected only with one of them. This state is strongly affected by the RG flow of the $\bar{g}_{\mathcal{B}R}$ coupling. If $\bar{g}_{\mathcal{B}R}$ flows to a non--zero fixed value, Eq.(\ref{rgeq5}) leads $\Delta k_{FR}$ to the QCR in this case. The $\bar{g}_{\mathcal{B}R}$ renormalization suffers the direct opposition of the umklapp interaction$\bar{g}_{\mathcal{U}R}$. If $\bar{g}_{\mathcal{B}R}$ flows to a non--zero fixed value this is followed immediately by $\bar{g}_{\mathcal{U}R}\rightarrow0$ and vice--versa. As a result, one can control the RG flow for the couplings by choosing appropriate initial values for $\bar{g}_{\mathcal{B}}^{ini}$ and $\bar{g}_{\mathcal{U}}^{ini}$. In this way, if we want to delay the $\Delta k_{FR}$ flow as much as possible we choose initial values for $\bar{g}_{\mathcal{B}}^{ini}$ and $\bar{g}_{\mathcal{U}}^{ini}$ in the weak coupling limit and this in turn produces a much slower renormalization of $\Delta k_{FR}$. For our purposes it suffices to choose $\Lambda_0/\Delta k_{F}^{ini}=10^3$ to ensure that the flow of $\ln\left(\Lambda_0/\Delta k_{FR}\right)$ remains at the upper region delimited by the $l=\ln\left(\Lambda_0/\Delta k_{FR}\right)$ line. Hence the flows not even touch this line even in a regime in which the quantum confinement is manifest, as can be seen in the small chart inside Fig. \ref{lamb0}(b).

\section{Numerical Results}

We solve all these coupled RG flow equations self-consistently making use of the fourth-order Runge-Kutta numerical method. The numerical stability of the flow equations depends on the Runge-Kutta step which in turn is directly related to our choices for both the RG energy scale $\omega$ and for the RG step $l$ parameters. In practice, we take $l_{max}=25$ and this gives $\omega/\Omega\sim10^{-11}$. 

If we sum directly the  RG flow equations for $\bar{g}_{0R}$ and $\bar{g}_{\mathcal{F}R}$ we arrive immediately at the no flow condition 

\begin{eqnarray}
\frac{d}{dl}(\bar{g}_{0R}(l)+\bar{g}_{\mathcal{F}R}(l))=0.
\label{g0ugf}
\end{eqnarray} 

\begin{figure}[b]
\centering
\includegraphics[scale=0.42]{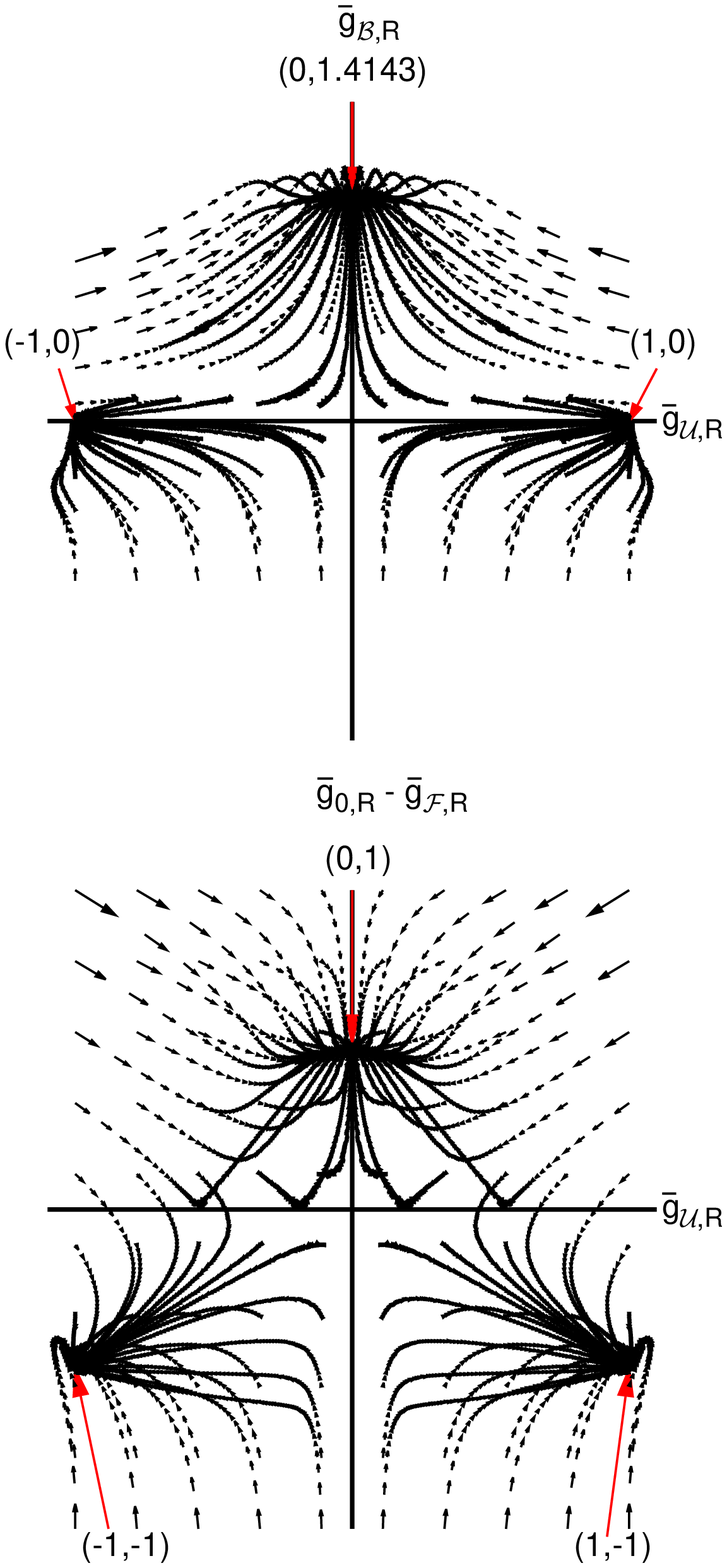}
\caption{(Color online) RG fixed points for the TCCM renormalized couplings.}
\label{conf2neg45}
\end{figure}

This result is particularly important in view of
the Luttinger theorem (LT) \cite{lutt,alt}. The LT relates the total Fermi wave vector $K_{FR}=k_{FR}^{a}+k_{FR}^{b}$ to the fermionic number 
density $n$. As a result if the number density is fixed, $K_{FR}$ does not flow, \emph{i.e.}, $\frac{dK_{FR}}{dl}=0$. 
Using this new condition together with the RG flow equations we immediately find that the LT is already trivially satisfied since indeed

{\small
\begin{eqnarray}
\frac{dk_{FR}}{dl}&=&-\frac{dk_{FR}^a}{dl}
=\nonumber \\
&-&\bar{g}_{\mathcal{B}R}^2\left[\frac{1}{1+\frac{\Delta k_{FR}}{\Lambda_0}e^l}+\frac{sign(1-\frac{\Delta k_{FR}}{\Lambda_0}e^l)}{\left|1-\frac{\Delta k_{FR}}{\Lambda_0}e^l\right|}\right].
\end{eqnarray}
}

In view of the LT there are only three independent renormalized couplings. For convenience, we choose to work with $\bar{g}_{0R}-\bar{g}_{\mathcal{F}R}$, $\bar{g}_{\mathcal{B}R}$ and $\bar{g}_{\mathcal{U}R}$ and we refer to this choice of couplings only from now on. Notice that our $\bar{g}_{0R}-\bar{g}_{\mathcal{F}R}$ is tantamount to the ``spin--like'' coupling defined in the Abelian bosonization scheme applied to a simpler two coupled chains model\cite{kn,nersesyan}.  

Although there are other fixed points present which result from the choice of different initial conditions we are directly interested only in two sets of them. They are determined basically by the two fixed values of $\bar{g}_{\mathcal{B}R}$, namely, $\bar{g}_{\mathcal{B}R}^*=0$ or $\bar{g}_{\mathcal{B}R}^*\approx\sqrt{2}$, the interchain backscattering coupling. As can be seen from our Eq. (\ref{rgeq5}), $\bar{g}_{\mathcal{B}R}$ plays a central role throughout the renormalization of $\Delta k_{FR}\propto t_{\perp}^{eff}$\cite{fabrizio,Dusuel,ledowski}. In view of that, $\Delta k_{FR}$ can suffer a strong renormalization which leads in practice to the coalescence of the two Fermi points. It turns out that indeed $\Delta k_{FR}\rightarrow0$ under special conditions. However, when this comes about the Luttinger liquid state suffers a crossover to yet another non--Fermi liquid state\cite{kn,fabrizio,Dusuel,nersesyan}. In other words, the QCR is accompanied by the opening of a charge gap and this in turn produces a new two--band structure in the single particle spectrum. That is, this non-zero fixed value of the interchain backscattering interaction is associated with the nullification of the transverse interchain hopping\cite{fabrizio,Dusuel,ledowski}, since $\Delta k_{FR}=t_{\perp}/2v_{FR}$ in our case. In contrast, as soon as $\bar{g}_{\mathcal{B}R}\rightarrow0$ the RG flow for $\Delta k_{FR}$ practically stops at a non--zero value. Notwithstanding that the quasiparticle weight $Z\rightarrow0$ and the anomalous dimension $\gamma^*$ becomes non--zero. In this case, there is no QCR but nevertheless there is a crossover from a Fermi liquid to a Luttinger liquid state as we show below.   

We display the flows for the renormalized couplings in Fig. \ref{conf2neg45}. The two sets of fixed points are directly influenced by our choice of initial conditions for the start up of the flows. The first set is characterized by $\bar{g}_{\mathcal{B}R}^*=1.4143\approx\sqrt{2}$,  $\bar{g}_{0R}^{*}-\bar{g}_{\mathcal{F}R}^{*}=1$ and $\bar{g}_{\mathcal{U}R}^*=0$. In contrast, the second set is tuned if the initial couplings are such that the renormalized couplings flow to $\bar{g}_{\mathcal{B}R}^*=0$,  $\bar{g}_{0R}^{*}-\bar{g}_{\mathcal{F}R}^{*}=-1$ and $\bar{g}_{\mathcal{U}R}^*=\pm 1$. If $\bar{g}_{\mathcal{U}R}=0$ from the beginning, the renormalized couplings flow to the fixed points much slower than if $\bar{g}_{\mathcal{U}}^{ini}\neq0$ at weak coupling. That is basically why the Luttinger liquid state stabilizes itself at weak--to--intermediate coupling, as pointed out by our analysis below. For initial coupling values of small magnitude these are the two relevant sets. We will discuss next the physical regimes associated with them.   

\begin{figure}[t]
\centering
\subfigure[]{\includegraphics[scale=0.30,angle=270]{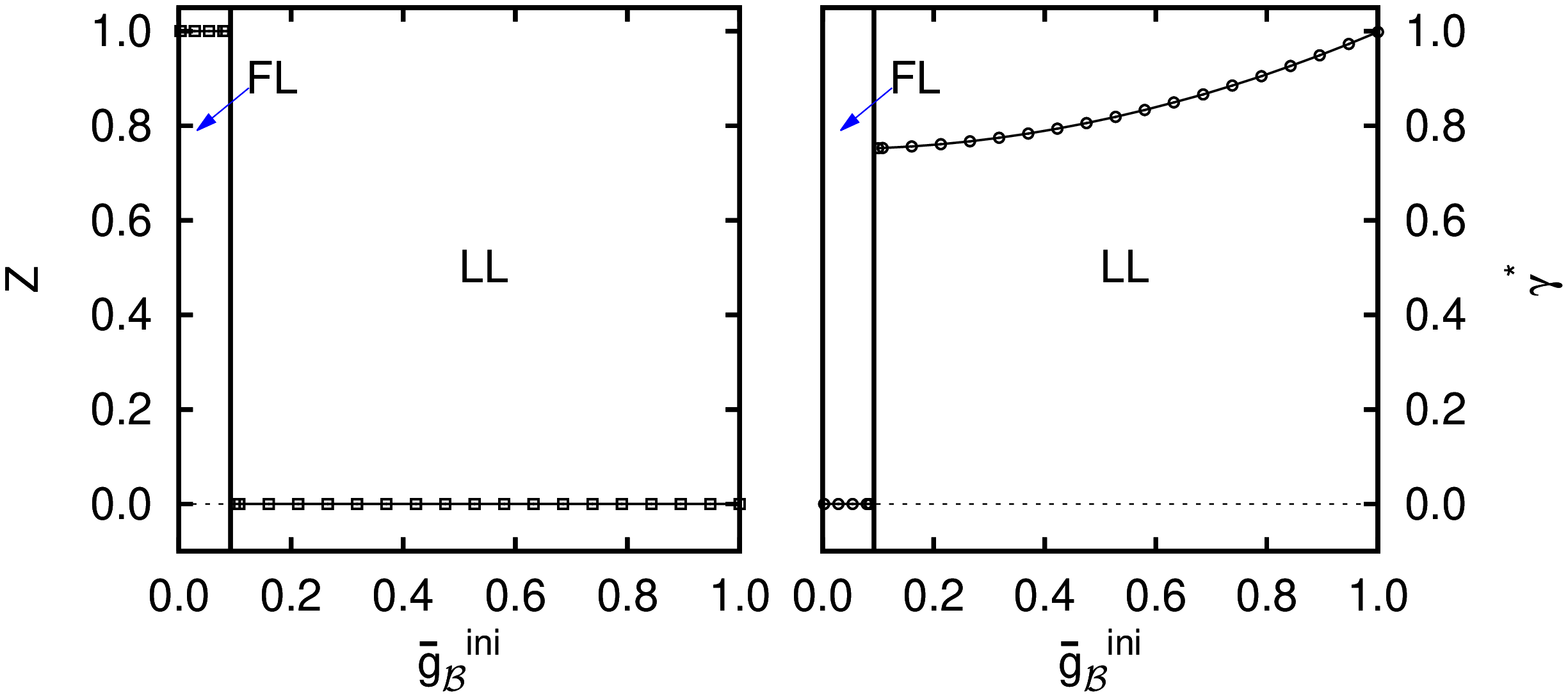}}
\subfigure[]{\includegraphics[scale=0.30,angle=270]{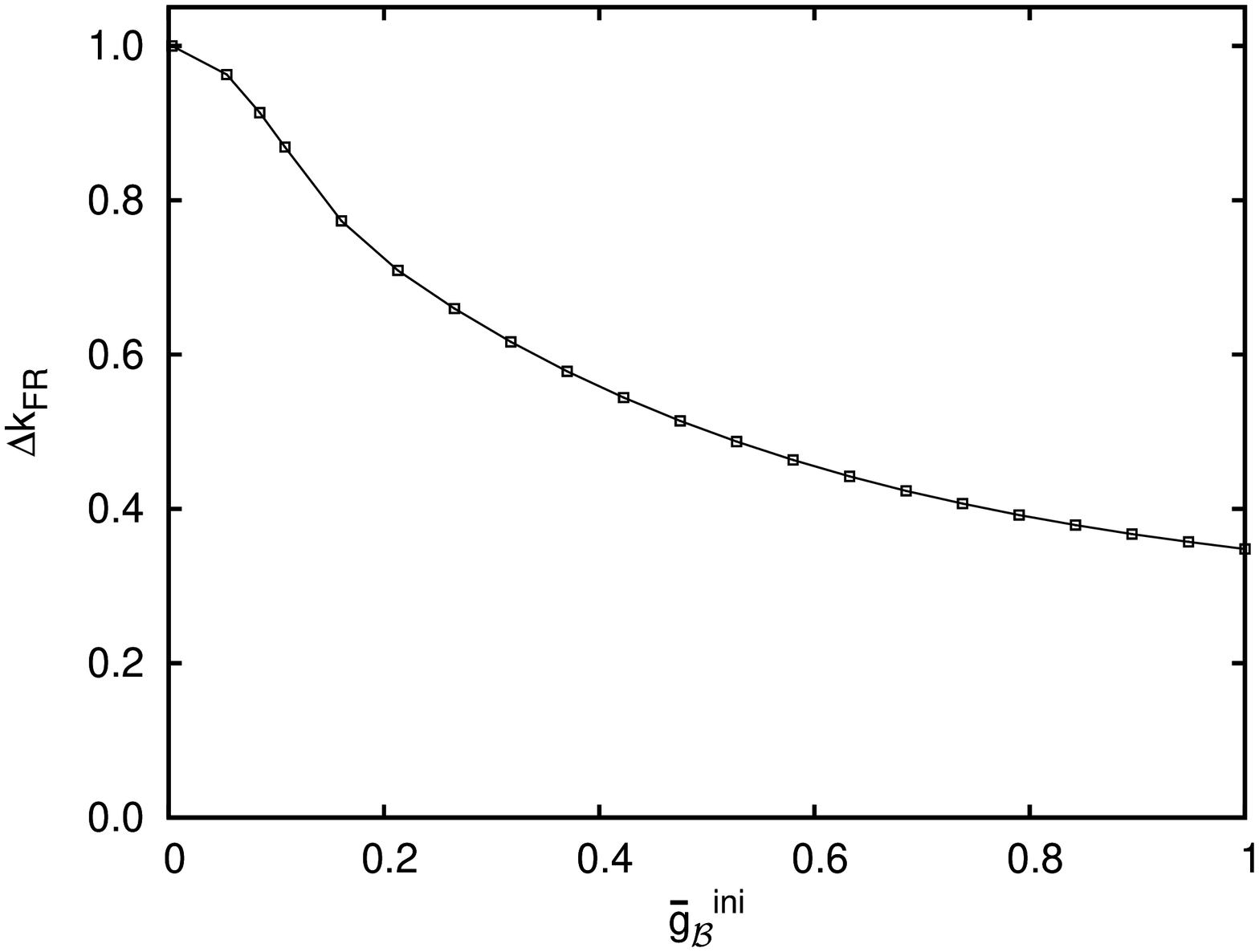}}
\caption{(Color online) (a) The quasiparticle weight $Z$ and anomalous dimension $\gamma^{*}$ as a function of $\bar{g}_{\mathcal{B}R}^{ini}$; (b) $\Delta k_{FR}$ as a function of the initial couplings $\bar{g}_{\mathcal{B}R}^{ini}$. Here, all couplings vary assuming that the initial coupling values are such that  $\bar{g}_{0}^{ini}-\bar{g}_{\mathcal{F}}^{ini}=-0.003$ and $\bar{g}_{\mathcal{B}}^{ini}=
\bar{g}_{\mathcal{U}}^{ini}$.}
\label{AD}
\end{figure}

\begin{figure}[t]
\centering
\includegraphics[scale=0.30,angle=270]{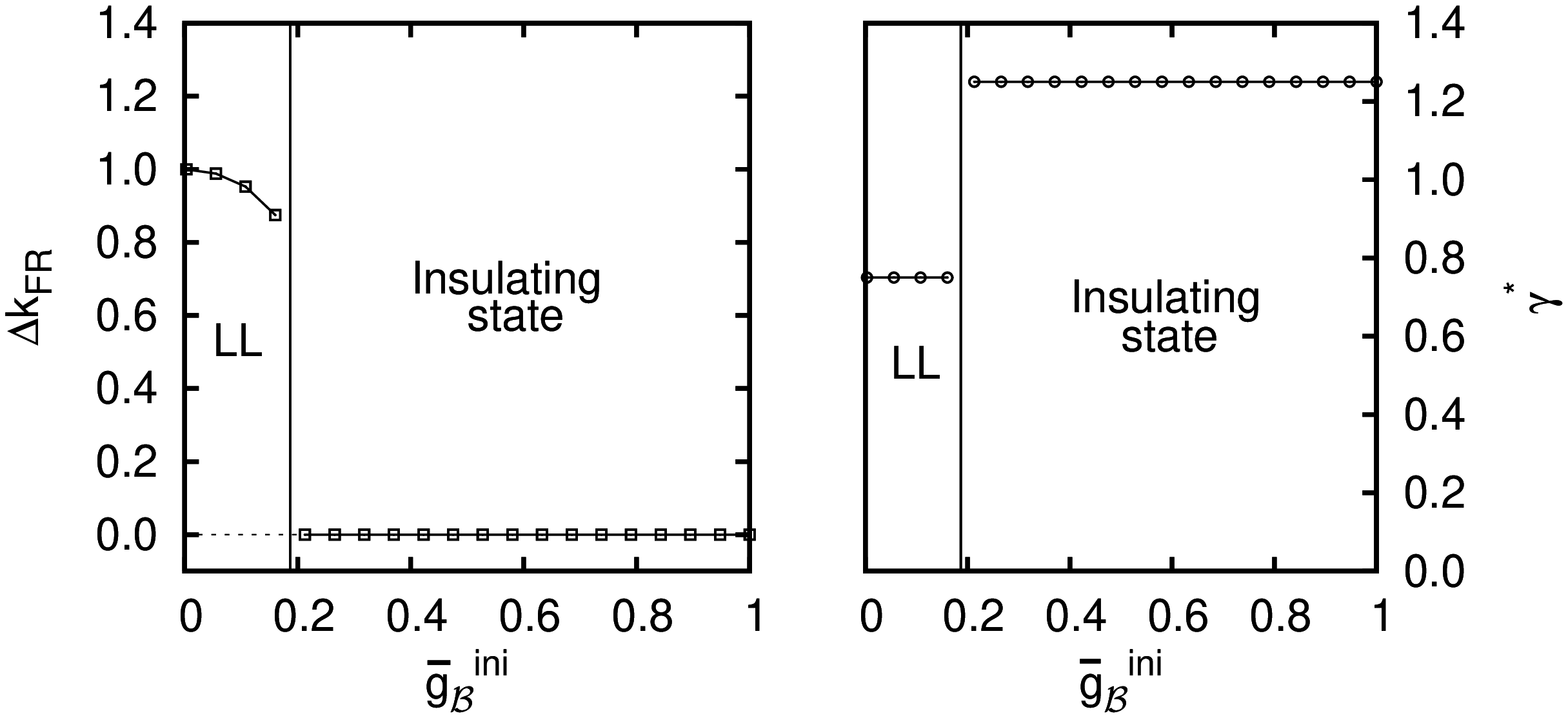}
\caption{Here we display the transition from a metallic NFL regime to an insulating fluid characterized by the QCR regime through $\Delta k_{FR}$ and the anomalous dimension $\gamma^{*}$ as a function of $\bar{g}_{\mathcal{B}}^{ini}$ for $\bar{g}_{0}^{ini}-\bar{g}_{\mathcal{F}}^{ini}=-0.1$, $\bar{g}_{\mathcal{U}}^{ini}=0.1$ as initial conditions.}
\label{conf2neg46}
\end{figure}

Let's move on to Figs. \ref{AD}(a), \ref{AD}(b) and \ref{conf2neg46}. If the renormalized couplings never reach fixed point values, the anomalous dimension
$\gamma^{*}=0$ and we are clearly dealing with a FL state in this case\cite{Dusuel}. For a nonzero $\gamma^{*}$, in the vicinity of the FPs, the single particle propagator scales as $G_{iR}^j(\omega,0)\approx\omega^{-(1-\gamma^{*})}$. The related occupation number $n_k^j$, at the Fermi points, scales with 

\begin{eqnarray}
\left.\frac{\partial n_k^j}{\partial k}\right|_{k_{FR}^j}\sim|k_{FR}^j-\alpha k|^{\gamma^{*}-1}
\end{eqnarray}

If the anomalous dimension is such that $0<\gamma^{*}<1$ there are no quasiparticle states and the single particle propagator displays a branch cut behavior. This is realized in our flows for $\bar{g}^{ini}\gtrsim0.1$. We find in this case $0.78\lesssim\gamma^{*}<1$, as shown in Fig. \ref{AD}(a). This metallic behavior is directly associated with the infinite slope of $n_k^\alpha$ at the Fermi points for these values of $\gamma^ {*}$. In contrast, if $\gamma^*>1$, $n_k$ is a smooth function of $k$ and, as a result, there is not even a residual Fermi surface left for the TCCM in this limit\cite{Dzy,mat,af}.      

We display these two regimes in our Fig. \ref{AD}(a). In Fig. \ref{AD}(b) we show how $\Delta k_{FR}$ evolves as we vary the initial values of the renormalized couplings. As it can be observed, $\Delta k_{FR}$ is never nullified in the LL state. The asymptotic behavior of $\Delta k_{FR}$ reveals that the quantum confinement regime (QCR) is never reached even for larger values of $\gamma^{*}$ which are no longer appropriate for a LL state. This occurs because we are varying the initial coupling values in such a way that the renormalized couplings always flow to the second set of fixed points, namely $\bar{g}_{0R}-\bar{g}_{\mathcal{F}R}\rightarrow-1$, $\bar{g}_{\mathcal{U}R}\rightarrow\pm1$ and $\bar{g}_{\mathcal{B}R}\rightarrow0$.

In Fig. \ref{AD}(a) we also display that the FL regime is stable for very low initial coupling values. This state stabilizes itself thanks to small initial values of umklapp interactions which act to delay the flow of all RG quantities towards the fixed points\cite{Dusuel}. In this regime, the couplings never approach a fixed point value. As a consequence of that, the quasiparticle weight $Z$ is never nullified and $\gamma$ never stabilizes for our choice of $l_{max}=25$. There is in reality no anomalous dimension in this case and $\Delta k_{FR}$ is practically left unchanged as it should be for a FL state. In fact, for greater initial values of $\bar{g}_{\mathcal{U}}^{ini}$ the LL state stabilizes and a transition from FL to LL state takes place for even lower values of $\bar{g}_{\mathcal{B}}^{ini}$. As a result, one can infer that the umklapp ``$g_2$--like'' interactions act to stabilize the LL state at weak--to--intermediate couplings bringing the transition region closer to the weak coupling regime. The situation is completely different when one considers the transition from the LL state to the QCR at intermediate--to--strong coupling regime. Small values of $\bar{g}_{\mathcal{U}}^{ini}$ acts to accelerate the flow to the fixed point values.      

In Fig. \ref{conf2neg46} we show how these flows are modified when the initial value of the interchain backscattering coupling is increased. We move on now to the domain of the first set of renormalized fixed points (\emph{i.e.}, $\bar{g}_{0R}-\bar{g}_{\mathcal{F}R}\rightarrow1$, $\bar{g}_{\mathcal{U}R}\rightarrow0$ and $\bar{g}_{\mathcal{B}R}\rightarrow1.4143\approx\sqrt{2}$). It is enough for us to vary only the interchain backscattering initial values with the other initial couplings being kept fixed ($\bar{g}_{0}^{ini}-\bar{g}_{\mathcal{F}}^{ini}=-0.1$ and $\bar{g}_{\mathcal{U}}^{ini}=0.1$) at weak coupling as initial conditions. For $\bar{g}_{\mathcal{B}}^{ini}\gtrsim0.2$, $\Delta k_{FR}$ suffers a discontinuity, $\gamma^{*}$ jumps accordingly to a larger value of $\gamma^{*}\sim1.23$ and the QCR becomes manifest in this new state. The presence of a $\gamma^{*}>1$ indicates that $n_k^j$ is smooth at the Fermi points and that there is no trace of the Fermi surface in this resulting state. In addition, the fact that $\bar{g}_{0R}-\bar{g}_{\mathcal{F}R}\rightarrow1$ to some extent reveals that the nature of this state can be directly associated with correlated charge density waves. This fact was evidenced in an earlier bosonization work considering a similar two--chains model without umklapp interactions\cite{nersesyan}. This charge gap in the single particle spectrum signalizes that the quantum confinement state is associated with a charge density wave insulator.      

\section{Conclusions}

We perform a self-consistent field theoretical renormalization group treatment of the two coupled chains model (TCCM) for spinless fermions. Making use of the appropriate prescriptions for the renormalized one-particle irreducible four-point functions $\Gamma^{j(4)}_{iR}(\lbrace{\bf{k}}_{i}\rbrace)$ ($i=0,\mathcal{F},\mathcal{B},\mathcal{U}$) and for the renormalized inverse fermionic single-particle propagator $\Gamma^{j(2)}_{\alpha R}({\bf{k}})$, we established the flow equations for the renormalized couplings, the quasiparticle weight and for the Fermi momenta difference $\Delta k_{FR}$. The Luttinger theorem naturally holds for all coupling regimes and help us to simplify and to write out our set of flow equations in a more concise form. It turns out that the difference of the bonding and antibonding Fermi points (FPs) $k_{FR}^{b}-k_{FR}^{a}=\Delta k_{FR}$ is only nullified at strong coupling for a non-zero fixed value of $\bar{g}_{\mathcal{B}R}^{*}\approx\sqrt{2}$. In our analysis we select two sets of intermediate to strong coupling fixed points regimes, which are tuned by our choice of initial conditions. In one of these two regimes there is a transition from a Fermi liquid state (FL)-characterized by a quasiparticle weight $Z\lesssim1$ for small initial values for the couplings-to the Luttinger liquid state (LL). This case is tuned by the second set of fixed points which is characterized by the condition $\bar{g}_{\mathcal{B}R}^*=0$. In spite of the fact that $Z=0$ we observe no quantum confinement regime (QCR) for intermediate initial coupling values. In contrast, if we choose initial conditions which lead us to the first set of intermediate--to--strong fixed points $(\bar{g}_{\mathcal{B}R}^*\approx\sqrt{2})$, there is never a FL state and the transition takes place between a LL and a charge density wave insulator NFL state\cite{kn,fabrizio,Dusuel,nersesyan} with $\gamma^*>1$. After such a transition we observe the QCR typified by the flow to zero of $\Delta k_{FR}$. This NFL with large $\gamma^*$ is characterized by a smooth momentum distribution function $n_k^j$ at the Fermi points which is an indicative of the presence of a charge gap in the single particle spectrum. In other words the QCR only takes place in a NFL which is already a charge density insulator. 

In conclusion, for small initial couplings--for which the interchain backscattering $(\bar{g}_{\mathcal{B}R})$ flows to zero--we have $Z \lesssim 1$, $\Delta k_{FR}$ barely changes from its initial value and the resulting state is a Fermi liquid. For larger initial couplings with a zero fixed point value for $\bar{g}_{\mathcal{B}R}$, the quasiparticle weight $Z$ is nullified, $\Delta k_{FR}$ remains non--zero and the resulting state is a LL. Early works\cite{kn,fabrizio,nersesyan} with a similar two--chains model without umklapp interactions pointed out that the physical system suffers a transition for a state in which there is a strong suppression of the electronic hopping between sites. The transverse hopping in this case is mediated by the flow of the interchain backscattering interaction to a new strong coupling fixed point. That is exactly what results from our RG analysis. For appropriate initial conditions $\bar{g}_{\mathcal{B}R}$ approaches a non--zero fixed point value, $\Delta k_{FR}$ becomes zero, signalizing the QCR, and the resulting state is a charge density wave insulator. Taking into account that $\bar{g}_{\mathcal{U}R}^*=0$, the TCCM most likely belongs to the same universality class of the half--filled one dimensional Hubbard model(1D--HM) in this limit. At half-filling the 1D--HM develops a charge gap and becomes a Mott insulator\cite{korepin}. The fact that $\bar{g}_{0R}-\bar{g}_{\mathcal{F}R}\rightarrow1$ in the QCR reinforces this possibility even further. For the 1D--HM $\gamma^*=1$. If we take higher loops corrections into account our estimate may be reduced from its critical $\gamma^*>1$ limit and approach this exact value. The presence of the umklapp interactions affect the RG flows in such a way that the QCR takes place only for larger initial values of $\bar{g}_{\mathcal{B}R}^{ini}$ when the RG flows for the remaining couplings approach the first set of fixed points. The opposite trend is observed when the FL state is stabilized in the weak coupling regime by the $\bar{g}_{\mathcal{U}R}$ approaching a non--zero value and by the backscattering coupling $\bar{g}_{\mathcal{B}R}$ flowing to zero\cite{Dusuel}. It is important to investigate the nature of the resulting QCR state further by employing other complementary non--perturbative tools. This work is under way and will be presented elsewhere.     

\begin{acknowledgement}
A. F. thanks the MCTI (Brazil) and the CNPq (Brazil) for financial support. We acknowledge the participation of Dr. T. Prudencio in the early stages of this work. One of us (AF) is grateful to Vladimir E. Korepin for discussions and for critically reading our manuscript.    
\end{acknowledgement}

\end{document}